\setuppapersize[letter][letter]
\setuplayout[topspace=0.5in,header=24pt,height=middle]
\setuppagenumbering[location={},style=bold]
\setupfooter[style=\it]

\setupcolors[state=start]
\definecolor[headingcolor][black]
\setuphead[section][style=bold,,color=headingcolor]
\setuphead[chapter][color=headingcolor, style={\ss\bfd},prefix=+]
\setuphead[section][color=headingcolor, style={\ss\bfc}]
\setuphead[subsection][color=headingcolor, style={\ss\bfa}]
\setuphead[title][style={\ss\bfd}]
\setuphead[subsubject][after={\vskip-\smallskipamount},before={\bigskip\bigskip}]

\setupwhitespace[medium]
\setupitemize[style={\sl}]

\def\rfig#1{\placefigure[right,none]{}{\externalfigure[#1]}}

\def\mydate{16 April 2007}
\def\id{0704.1854v1}
\def\arxiv{arXiv:\id\ [physics.ed-ph]\quad \mydate}

\definelayer	[arxiv]  
		[width=\paperwidth, height=\paperheight]
\relax
\setlayer	[arxiv]	
		[x=0.5in, y=3in]
		{\setupbodyfont[14pt]{\rotate[rotation=90]{\ss
\arxiv}}}

\setupbackgrounds	[page]	[background=arxiv]

\starttext

\title{Teaching for transfer}

\start
\rightskip=0pt plus 1fill\obeylines
\setupwhitespace[1.5pt]
{\it Sanjoy Mahajan}
\vskip3pt
Teaching and Learning Laboratory
\quad \&\ Dept of Electrical Engineering and Computer Science
MIT, Room 5-122
Cambridge, MA 02139
{\tt sanjoy@mit.edu}
\bigskip
\mydate
\stop
\vskip1in

  {\bf Abstract.}  Students, after they leave our care, are called to
  solve the diverse problems offered by the world, so we should teach to
  increase {\it transfer\/}: the ability to apply fundamental principles to
  new problems and contexts.  This ability is rare.

  The following pages are from a workshop for faculty on designing
  courses that promote transfer.  I discuss two design principles: to
  {\it name\/} the transferable ideas and to illustrate them with
  examples from {\it diverse subjects}.  The discussion uses
  dimensional reasoning as the example of a valuable transferable
  idea, illustrating it with three diverse examples.

\vfill
\start \switchtobodyfont[9pt]

  Copyright 2007 Sanjoy Mahajan.  This document is {\it free
    software}.  You can modify and/or redistribute it under the terms
  of the GNU General Public License as published by the Free Software
  Foundation: either version 2 of the License or (at your option) any
  later version.  The source code is available at \hbox{\tt
  arxiv.org/abs/\id} from the 'Other formats' link.
\par
\stop

\page
\setuppagenumbering[location={header,right},style=bold]
\setupfootertexts[Licensed under the GNU GPL v2 or later\hfill]

\def\1#1{{\it [#1]}}

\subject{One theme to bind them all}

\1{I don't give away the name of the technique yet, hence the coy
  section title, so that participants can discover it by discussion.}

\subsubject{Pyramid volume}

\1{I ask the participants to discuss, in small groups, the flaws in
  the proposed volumes.  Then we name the technique -- dimensional
  analysis -- for use in the next examples.}

Here is a pyramid with a square base.  
It has height $h$ and the side
of the base is $b$.  Comment on these proposed formulas for the volume:
\startitemize[a]
\def\1#1{\item$\displaystyle{#1}$}
\1{{1\over 3}bh}
\1{{h^3+b^2+b}}
\1{{b^4+h^3}}
\1{{b^4/h}}
\stopitemize

\vfil
\subsubject{GDP}

\1{To teach for transfer, people must practice transfer to other
  domains.  So I ask the audience to find the worst flaw in the
  following argument, which comes from a domain far from physics or
  mathematics.}

In many articles criticizing globalization (the kindler, gentler name for
imperialism), you can read an argument like this one
[from `Impunity for Multinationals',\hfil\break
{\tt www.globalpolicy.org/socecon/tncs/2002/0911impunity.htm},
11 Sept 2002]:
\startnarrower \it
In Nigeria, a relatively economically strong country, the GDP is \$99
billion. The net worth of Exxon is \$119 billion.  `When
multinationals have a net worth higher than the GDP of the country in
which they operate, what kind of power relationship are we talking
about?' asks Laura Morosini.
\stopnarrower

Find the most egregious fault in this argument.

\vfil
\subsubject{Tidal waves}

\1{And a hard example of dimensional analysis, to show that the wave
  speed is $v=\sqrt{gh}$.}

The speed of water waves in shallow water depends on the depth and on
gravity, which provides the force that drives the waves.  Find a
formula that connects the speed $v$ to the gravitational acceleration
$g$ and to the depth $h$.

Tidal waves on the ocean are an example of shallow-water waves (!).
How fast do they travel?

\page

\subject{Why use diverse examples}

Here is a line of reasoning showing why diverse examples promote
transfer.

\subsubject{One example}

\rfig{fig.1}

A bare concept is difficult to grasp without examples.  Examples help
learners to understand the concept and, if chosen well, to separate the
transferable concept from the illustrations.  
Suppose then that you explain a concept and illustrate it with an
example.  The concept and example merge in the learner's mind, leaving
him or her uncertain about the boundaries of the concept.

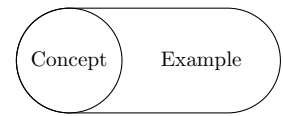
\subsubject{Two examples}

\rfig{fig.2}

One remedy is to offer a second example.  To the extent that the
second example is similar to the first, the first concept plus example
overlaps the second concept plus example.  The overlap includes a
penumbra around the concept.  The penumbra is smaller than when only
one example is offered, and the two examples delimit the boundaries of
the concept more clearly than when only one example is offered.
Progress!

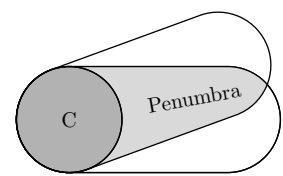
\subsubject{Diverse examples}

\rfig{fig.3}

You can help the learner even more by taking the second (or third)
example from a distant field: wherefore {\bf curriculum integration}.
The penumbra shrinks and the concept stands out.

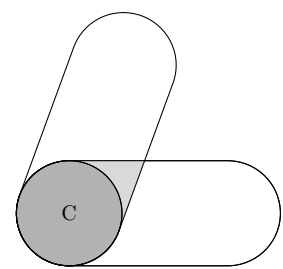
\vfill
{\bsb \baselineskip=2.6ex%
\noindent
\llap{`}Only connect! That was the whole of her sermon\dots Live in fragments no
longer\rlap{.'} \hfill\qquad\tfb E.$\,$M.$\,$Forster, {\sl Howard's End}\rlap{.}\par}
\bigskip

\stoptext